\documentclass{appolb}
\usepackage{epsfig}
\usepackage{graphicx}
\graphicspath{{PNLO/}{BHCPv3/}}
\usepackage{amssymb}
\usepackage{amsmath}
\usepackage{bm}
\usepackage{multirow}
\usepackage[normalem]{ulem}  % \sout{old text} for strikeout
\usepackage[dvips]{color} % For blue in-text comments and additions

\renewcommand\sout{\bgroup \color{red} \ULdepth=-.5ex \ULset}

\newcommand{\Comments}[1]{}
\newcommand{\nn}{\nonumber}

\newcommand{\Feff}{\mathcal{F}_\mathrm{eff}}

\newcommand{\ZNc}{Z_{N_c}}

\newcommand{\TQ}{T_\mathrm{Q}}
\newcommand{\PSC}{P-SC-LQCD}
\newcommand{\Psfig}[2]{\includegraphics[width=#1]{#2}}
\newcommand{\Msun}{M_{\odot}}
\newcommand{\muB}{\mu_{B}}

\newcommand{\rhoB}{\rho_{B}}
\newcommand{\MeV}{\mathrm{MeV}}
\newcommand{\fm}{\mathrm{fm}}
\newcommand{\PNJLx}{P-NJL$_8$}
\newcommand{\TCP}{T_\mathrm{CP}}
\newcommand{\muCP}{\mu_\mathrm{CP}}
\begin{document}
\title{
QCD critical point in the strong coupling lattice QCD
\\
and during black hole formation
%The Paper Title comes Here...%
\thanks{Presented at the HIC for FAIR workshop \& 28th Max Born Symposium
on Three days on Quarkyonic Island, Wroclaw, Poland, May 19-21, 2011.\\
Report No.: YITP-11-107}%
}
\author{A. Ohnishi$^1$, K. Miura$^2$, T. Z. Nakano$^{1,3}$, N. Kawamoto$^4$,\\
H. Ueda$^3$, M. Ruggieri$^{1,5}$, K. Sumiyoshi$^6$
%Put here the name(s) of the Author(s)
\address{$1.$ Yukawa Institute for Theoretical Physics, Kyoto University,\\
Kyoto 606-8502, Japan\\
$2.$ INFN Laboratori Nazionali de Frascati\\
%, I-00044 Frascati (RM), Italy}
$3.$ Department of Physics, Faculty of Science, Kyoto University\\
%,Kyoto 606-8502, Japan}
$4.$ Department of Physics, Faculty of Science, Hokkaido University\\
%, Sapporo 060-0810, Japan}
$5.$ Department of Physics and Astronomy, University of Catania\\
%, Via S. Sofia 64, I-95125 Catania, Italy}
$6.$ Numazu College of Technology
%, Ooka 3600, Numazu, Shizuoka 410-8501, Japan}
}
%\and
%the Name(s) of other Author(s)
%\address{and their affiliation}
}
\maketitle
\begin{abstract}
%Here comes the abstract
We discuss the QCD phase diagram from two different point of view.
%One of them is the relation of the chiral and $\ZNc$ deconfinement
%transitions.
We first investigate the phase diagram structure in the strong coupling
lattice QCD with Polyakov loop effects, and show that
the the chiral and $\ZNc$ deconfinement transition
boundaries deviate at finite $\mu$ as suggested from large $N_c$ arguments.
Next we discuss the possibility to probe the QCD critical point
during prompt black hole formation processes.
The thermodynamical evolution during the black hole formation
would result in quark matter formation,
and the critical point in isospin asymmetric matter may be swept.
$(T,\muB)$ region probed in heavy-ion collisions
and the black hole formation processes
covers most of the critical point locations
predicted in recent lattice Monte-Carlo simulations
and chiral effective models.
\end{abstract}
%12.38.Gc	Lattice QCD calculations (see also 11.15.Ha Lattice gauge theory)
%12.38.-t	Quantum chromodynamics (for quarks, gluons, and QCD in nuclear reactions, see 24.85.+p)
%12.39.Fe	Chiral Lagrangians
%11.10.Wx	Finite-temperature field theory
%11.15.Me	Strong-coupling expansions
%21.65.Qr	Quark matter (see also 12.38.Mh Quark-gluon plasma in quantum chromodynamics; 25.75.Nq Quark deconfinement, quark-gluon plasma production and phase transitions in relativistic heavy-ion collisions)
%25.75.Nq	Quark deconfinement, quark-gluon plasma production, and phase transitions (see also 12.38.Mh Quark-gluon plasma in quantum chromodynamics; 21.65.Qr Quark matter in nuclear matter)
%21.65.Mn	Equations of state of nuclear matter (see also 26.60.Kp Equations of state of neutron-star matter)
%26.50.+x	Nuclear physics aspects of novae, supernovae, and other explosive environments 
  
\section{Introduction}
\label{Sec:Intro}

QCD phase transition at finite temperature ($T$) and chemical potential
($\mu$) is attracting much attention in recent years.
The beam energy and system size scan programs at RHIC~\cite{BES}
and SPS~\cite{NA61} are running
to discover the critical point and the first order transition 
at finite $T$ and $\mu$,
and the discovery of the two solar mass neutron star~\cite{Demorest}
gives us some hints
on the phase transition at finite density.
Recent large $N_c$ arguments suggest the existence of another form of matter,
referred to as the quarkyonic matter, where the Polyakov loop is suppressed
and the density is high~\cite{Quarkyonic}.

In this proceedings, we first discuss the chiral and $\ZNc$ deconfinement
transitions at finite $T$ and $\mu$ based on the strong coupling lattice
QCD with Polyakov loop effects (\PSC)~\cite{PSC}.
Next we discuss the possibility to probe the critical point
in isospin asymmetric high baryon density matter
formed during the prompt black hole formation processes~\cite{BHCP}.

\section{Strong coupling lattice QCD and quarkyonic matter}
\label{Sec:PSC}

Do the chiral and $\ZNc$ deconfinement transition boundaries
deviate at large $\mu$ ? 
This is one of the most interesting questions
in the current QCD phase diagram studies.
From the large $N_c$ arguments, we expect the existence of the so-called
quarkyonic matter, where the $\ZNc$ order parameter (Polyakov loop)
is suppressed and the density is high~\cite{Quarkyonic}.
%
%Some effective models support the existence of the quarkyonic matter,
%while others do not.
%Lattice QCD Monte-Carlo simulation has the sign problem at finite $\mu$
%and precise results are not yet obtained,
In the lattice QCD Monte-Carlo simulations at $\mu=0$,
%One of the interesting observations in the LQCD-MC is that
the $\ZNc$ transition temperature ($T_c(\ZNc)$)
is close to but somewhat larger than the chiral transition temperature
($T_c(\chi)$)~\cite{BW2006}.
In chiral effective models,
some of them predict the existence of quarkyonic-like
matter~\cite{PNJL-A,Tuominen},
% with small Polyakov loop and high density
while some of them predict that the $\ZNc$ transition boundary
agrees with the chiral transition
boundary~\cite{PNJL-B}.
Thus it is important to discuss the chiral and $\ZNc$ transition
boundaries at finite $\mu$ in the theoretical framework
directly based on QCD for $N_c=3$.

%==========================
\begin{figure}[b]
\includegraphics[width=6.5cm]{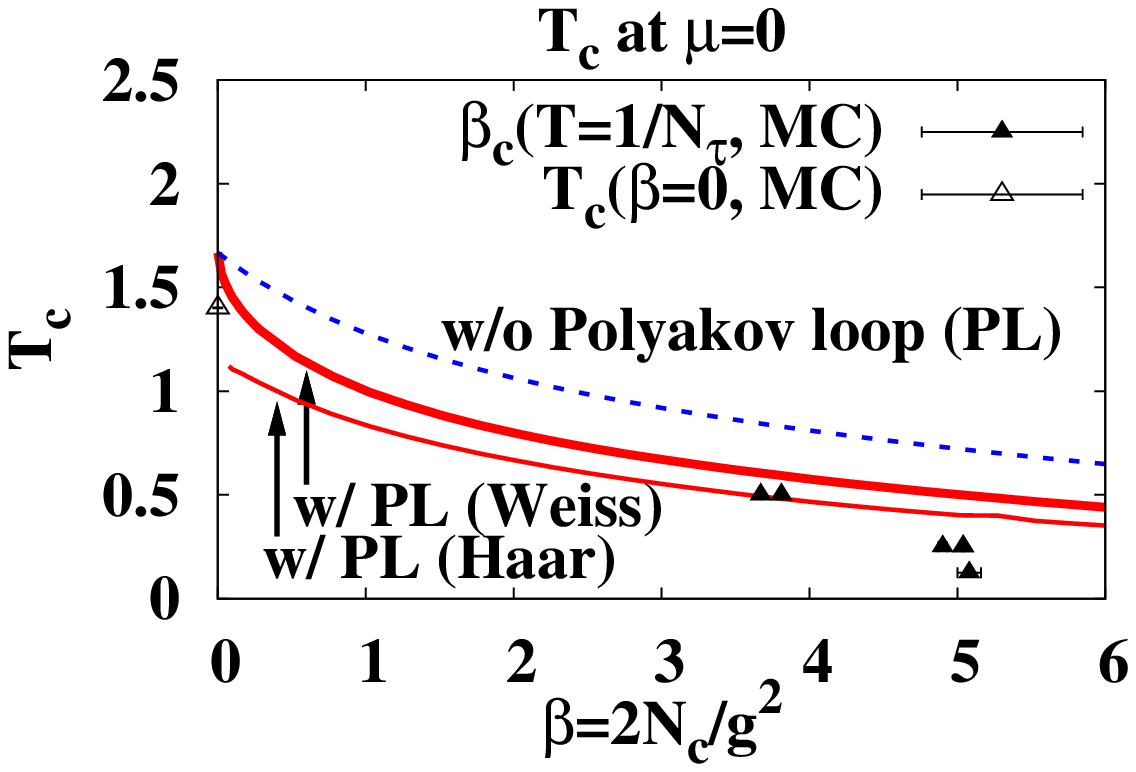}~\includegraphics[width=6.0cm]{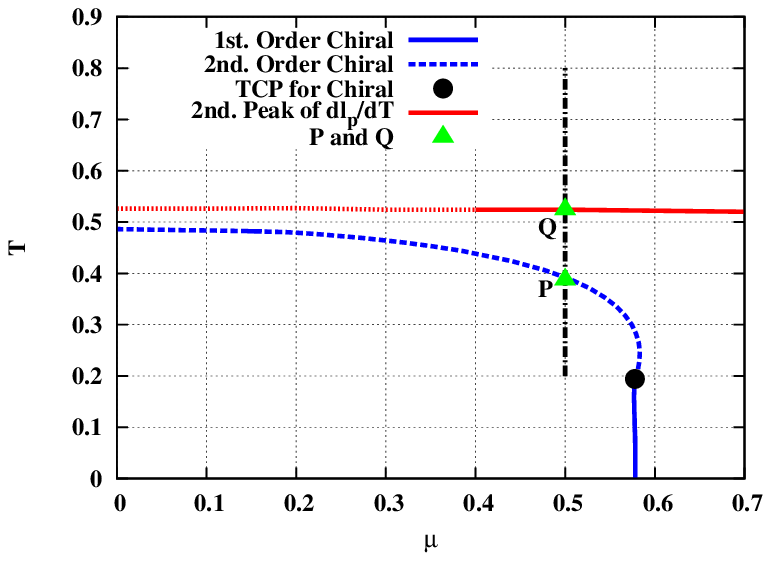}
\caption{
Left: Comparison of $T_c$ at $\mu=0$ in P-SC-LQCD (solid) 
and SC-LQCD (without the Polyakov loop, dashed)~\protect\cite{PSC}.
%in the lattice unit
Two different Polyakov loop treatments (Weiss and Haar method) are compared.
The triangles represent the lattice MC results
with one species of unrooted staggered fermion.
Right:
The phase boundary for the chiral transition
with the second peak of $d\ell_p/dT$
at $\beta=4$ in the chiral limit.
The ``P'' and ``Q'' correspond to
those in Fig.~\protect\ref{Fig:SigPol}.
}
\label{Fig:PD}
\end{figure}
%==========================

The strong coupling ($1/g^2$) expansion in the lattice QCD (SC-LQCD)
has been successful since the beginning of the lattice gauge
theory~\cite{SCLQCD-YM}
and would provide an alternative lattice framework to study
the QCD phase diagram at finite $T$ and $\mu$~\cite{Langelage,SCLQCD-Q}.
For example, the spontaneous breaking of the chiral symmetry
and its restoration at finite $T$ and/or $\mu$ have been known to be
realized in the strong coupling limit~\cite{SCLQCD-Q}.
Recently, the finite couping and Polyakov loop effects 
are incorporated in the framework of SC-LQCD,
and are found to explain the MC results of $T_c$ at $\mu=0$
in the region $\beta=2N_c/g^2 \lesssim 4$~\cite{PSC,FiniteBeta},
as shown in the left panel of Fig.~\ref{Fig:PD}.

We discuss here the chiral and $Z_{N_c}$ deconfinement dynamics
by using the SC-LQCD with the Polyakov loop effects (\PSC) 
in the mean field approximation~\cite{PSC}.
We take account of the next-to-leading order (NLO, ${\cal O}(1/g^2)$)
and the leading order ($\mathcal{O}(1/g^{2N_\tau})$)
of the strong coupling expansion in the fermionic and pure gluonic sector,
respectively, and in the leading order of the $1/d$ expansion~\cite{LargeD}.
The effective potential is given as~\cite{PSC}
%==========================
\begin{align}
&\Feff(\Phi;T,\mu)
\equiv -(T\log \mathcal{Z}_{\mathrm{LQCD}})/N_s^d
=\Feff^{\chi}+\Feff^\mathrm{Pol}
\ ,\label{eq:Feff}
\\
&\Feff^\chi
\simeq
\left(
\frac{d}{4N_c}+\beta_s\varphi_s
\right)\sigma^2
+\frac{\beta_s\varphi_s^2}{2}
+\frac{\beta_{\tau}}{2}
\bigl(\varphi_{\tau}^2-\omega_{\tau}^2\bigr)
-N_c\log \sqrt{Z_+Z_-}
%\Zchi=\sqrt{Z_+Z_-}\ ,
\nn\\
&~~~~~~~~
-N_cE_q
-T\left[\log\mathcal{R}(E_q-\tilde{\mu},\ell,\bar{\ell})
       +\log\mathcal{R}(E_q+\tilde{\mu},\bar{\ell},\ell)\right]
\ ,\label{eq:FeffChi}
\\
&\Feff^\mathrm{Pol}\simeq
-2TdN_c^2\biggl(\frac{1}{g^2N_c}\biggr)^{1/T}\bar{\ell}_{p}\ell_{p}
-T\log\mathcal{M}_{\mathrm{Haar}}(\ell_p,\bar{\ell}_p)
\ ,\label{eq:FeffPol}
\\
&
%\Zchi=\sqrt{Z_+Z_-}\ ,
\tilde{\mu}=\mu-\log\sqrt{Z_+/Z_-}\ ,\quad
Z_{\pm}=1+\beta_{\tau}(\varphi_\tau\pm\omega_\tau)\ ,
\nn\\
&\mathcal{R}(x,\ell,\bar{\ell})
\equiv
1
+N_c\ell e^{-x/T}
+N_c\bar{\ell} e^{-2x/T}
+e^{-3x/T}
\ ,\label{eq:Dq}\\
&\mathcal{M}_{\mathrm{Haar}}(\ell,\bar{\ell})
%\nn\\
%&\quad 
=1-6\bar{\ell} \ell
-3(\bar{\ell}\ell)^2
+4(\ell^3+\bar{\ell}^3)
\ ,
\end{align}
%==========================
where
$\sigma$ and $\ell$($\bar{\ell}$) denote
the chiral condensate and the (anti-)Polyakov loop, respectively,
$d=3$ is the spatial dimension,
$\beta_{\tau}=\beta d/2N_c^3$ and $\beta_s=\beta d(d-1)/16N_c^5$.
Several other auxiliary fields ($\varphi_s, \varphi_\tau, \omega_\tau$) are
introduced in addition to $\sigma$ and $\ell$,
and the stationary conditions are imposed on these fields in equilibrium.
The Polyakov loop $\ell$ couples with quarks,
and appears with the Boltzmann factor $e^{-(E_q-\tilde{\mu})/T}$.
Color-singlet states dominate in the confined phase ($\ell \sim 0$),
while quarks can excite in the deconfined phase ($\ell \neq 0$).
This point has been pointed out in the strong coupling limit
and utilized in the PNJL model~\cite{Fukushima2003}.

%==========================
\begin{figure}[tb]
\includegraphics[width=6.0cm]{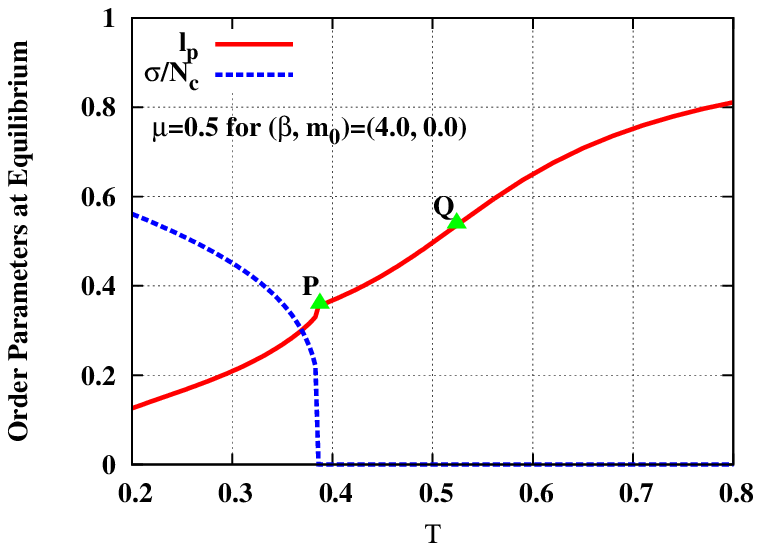}%
~\includegraphics[width=6.0cm]{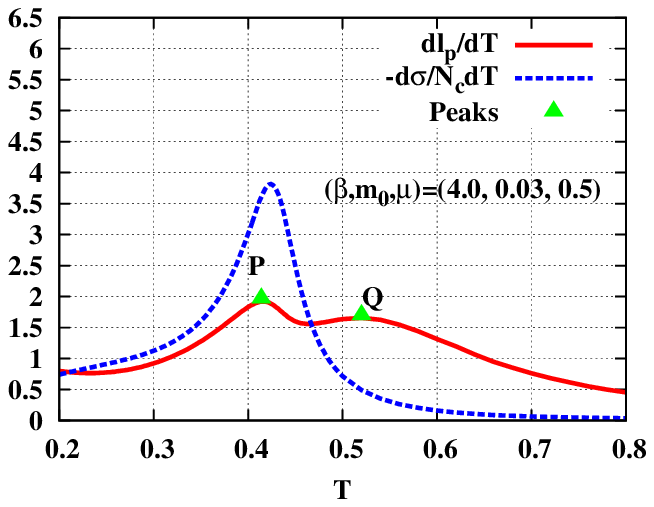}
\caption{
Left: $T$ dependence of
the chiral condensate (dashed) and the Polyakov loop (solid)
at $\mu=0.5$ and $\beta=4$ in the chiral limit.
"P" and "Q" correspond to the points in the right panel of Fig.~\protect{\ref{Fig:PD}}.
Right: $T$ dependence of $d\sigma/dT$ (dashed) and $d\ell_p/dT$ (solid)
with $m_0=0.03$ at $\beta=4$ and $\mu=0.5$.
%The values are shown in the lattice unit, $a=1$.
}
\label{Fig:SigPol}
\end{figure}
%==========================

%=========================================================
% Results
%=========================================================
We discuss the results at $\beta=4$,
which is in the coupling range where
\PSC\ roughly explains the LQCD-MC results of $T_c$ at $\mu=0$,
and results are shown in the lattice unit, $a=1$.
In the right panel of Fig.~\ref{Fig:PD},
we show the phase diagram at $\beta=4$ in the chiral limit ($m_0=0$).
%$T$ and $\mu$ are given in the lattice unit.
We find the second-(dashed) and first-order (solid)
chiral transition boundaries separated by
the (tri-)critical point (CP)
at $(\mu_{\mathrm{CP}},T_{\mathrm{CP}})=(0.58,0.19)$.

%We can also discuss the $\ZNc$ transition in \PSC.
In the left panel of Fig.~\ref{Fig:SigPol},
we show the $T$ dependence of the chiral condensate $\sigma$
and the Polyakov loop $\ell_p$ at $\mu=0.5$.
At the chiral transition temperature,
the chiral condensate becomes zero quickly,
and the Polyakov loop is affected to have a kink.
In addition to this chiral-induced kink shown as "P" in the figure,
we find another temperature shown as "Q"
where the Polyakov loop increases rapidly.
This feature is more clearly seen in $d\ell_p/dT$.
We show $d\ell_p/dT$ at $\mu=0.5$ and $m_0=0.03$
in the right panel of Fig.~\ref{Fig:SigPol}.
We can see the second peak in $d\ell_p/dT$ at $\TQ \sim 0.52$.
A similar double-peak structure has been reported
in the model studies based on PNJL model~\cite{Tuominen}.

The peak "Q" can be understood as the $\ZNc$-induced crossover
from following reasons.
The temperature $\TQ \sim 0.52$ is found to be almost independent
on the chemical potential, as indicated by the upper line in 
the right panel of Fig.~\ref{Fig:PD}.
$\TQ$ is also insensitive to quark mass.
The temperature of the peak ``P'' is shifted upward and becomes closer to ``Q''
with increasing $m_0$, while $\TQ$ stays almost constant.
For larger masses, $m_0>0.05$, the two peaks merges to one,
%a single peak,
and it grows with increasing $m_0$ at nearly $m_0$ independent temperature.
%and its temperature is nearly $m_0$ independent, $T\sim0.52$.
%
These observations agrees with the expected character
of the $\ZNc$-induced transition;
the $\ZNc$ deconfinement transition nature would be stronger
with large quark mass,
and its transition temperature would have small dependence
on $\mu$ from the large $N_c$ argument.
This interpretation supports the existence
of the quarkyonic-like matter in cold-dense matter,
where the Polyakov loop is suppressed and density is high.

\section{Critical Point Sweep during Black Hole Formation}
\label{Sec:BHCP}

Critical point (CP) is one of the largest targets
in the beam energy and system size scan programs at RHIC and SPS
and in the forthcoming FAIR facility.
CP is expected to be probed in these experiments
if it is in the low baryon chemical potential region,
$\muCP \lesssim 500~\MeV$,
as predicted in some of the lattice MC calculations~\cite{LR04,LT04,LC11}.
On the other hand, we cannot reject the possibility
that the CP is located in the lower $T$ and higher $\mu$ region,
as predicted in many of the chiral effective models~\cite{NJL,Fukushima2003,PNJL,PNJLx,PQM}.
Therefore, it is important to search for other candidate sites 
where hot and dense matter is formed and CP is reachable.

%------------------------------------------------------------------------*
\begin{figure*}[bt]
\begin{center}
\Psfig{12.5cm}{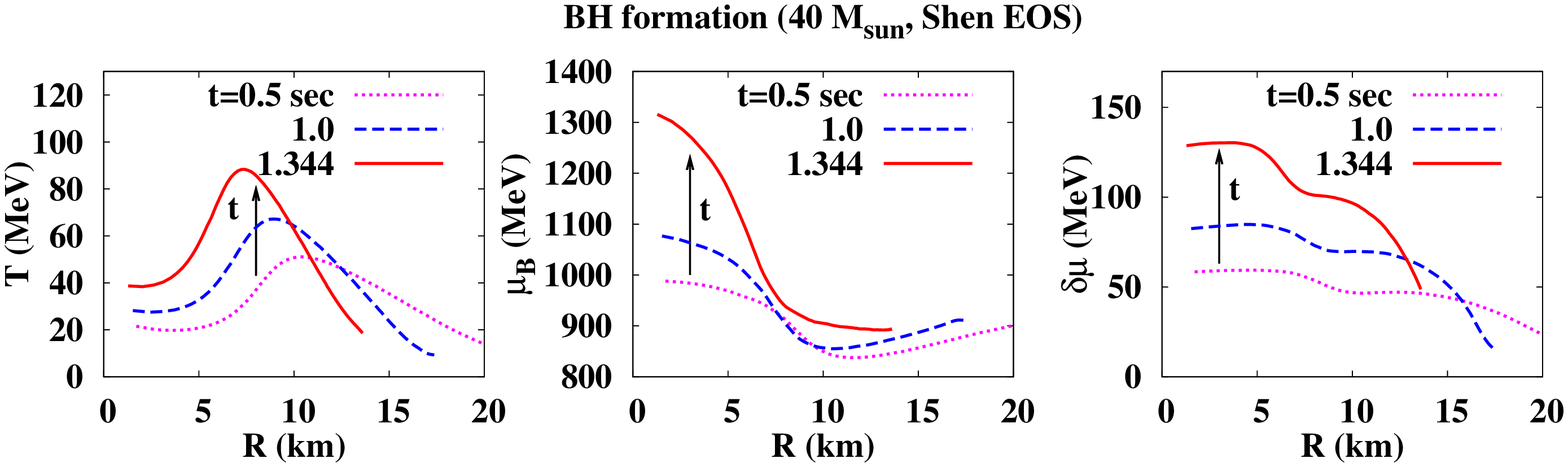}
\end{center}
\caption{
The BH formation profile, ($T, \muB, \delta\mu$), as a function of the radius.
%We also show the baryon density ($\rhoB$) and the electron fraction
%($Y_e \equiv \rho_e / \rhoB$).
Results are shown for the gravitational collapse of a 40 $\Msun$ star
at $t=$ 0.5 sec (dotted),
1.0 sec (dashed),
and 1.344 sec (solid lines, just before the horizon formation).
}\label{Fig:BH}
\end{figure*}
%------------------------------------------------------------------------*

A gravitational collapse of a massive star is a promising candidate.
%as the CP hunting site.
%
A majority of non-rotating massive stars with mass $M \gtrsim 20 \Msun$
are expected to collapse quietly (faint-supernova)
to black holes (BH)~\cite{StarFate}.
Their frequency should be comparable to supernovae
provided that the mass spectrum of stars has a long tail
as in the power law behavior.
The BH formation processes are found to form 
hot $(T \sim 90~\MeV)$ and dense $(\rhoB \sim 4 \rho_0)$ matter
in the neutrino-radiation hydrodynamical simulations
in the collapse and bounce stage of a 40$\Msun$ star~\cite{BH}.
Thermodynamical variables at a given time vary as a function of radius
in a proto-neutron star and form a line 
(referred to as the BH formation profile in the later discussions)
in the $T\mathrm{-}\muB$ plane.
%------------------------------------------------------------------------*
In Fig. \ref{Fig:BH}, we show the BH formation profile
$(T,\muB,\delta\mu)$~\cite{BH} calculated by using the Shen EOS
at $t=0.5, 1.0$ and $1.344$ sec after the bounce during the BH formation
from a 40 $\Msun$ star in the proto-neutron star core,
where the mass coordinate from the center is $M < 1.6 M_\odot$.
The time $t=1.344$ sec is just before the horizon formation.
From the outer to the inner region of the proto-neutron star,
$T$ first increases from $T\sim 10~\MeV$ to $T\sim~(50-90)~\MeV$
in the middle heated region, and decreases again inside.
The baryon chemical potential $\muB$ is found to go over 1300 MeV
in the central region just before the horizon formation at $t=1.344$ sec.
The isospin chemical potential is found to be $\delta\mu=(50-130)$ MeV
in the inner region.
%------------------------------------------------------------------------*
%
The temperature and density in BH formation are significantly higher than
% those
in the model explosion calculation of supernova.
The highest temperature and density are moderate in the collapse and bounce
stage of supernovae, $(T,\rhoB) \sim (21.5~\MeV, 0.24~\fm^{-3})$
when hadronic EOS is adopted~\cite{IOTSY},
while it has been argued that the transition to quark matter might trigger
successful supernovae~\cite{SNQ}.
%
%In hotter and denser environment during BH formation
%compared with the supernova explosions,
%we have a larger possibility of creating a new form of matter,
%such as the dense quark matter.

We discuss here the possibility that the BH formation profile evolves
with time and may pass through CP and the vicinity ({\em CP sweep}).
The CP location scatters in the $T\mathrm{-}\muB$ plane
in chiral effective models such as
NJL~\cite{NJL}, P-NJL~\cite{Fukushima2003,PNJL},
P-NJL with 8-quark interaction (\PNJLx)~\cite{PNJLx},
and PQM~\cite{PQM} models.
We expect that CP moves in the lower $T$ direction at finite $\delta\mu$,
because $d$-quark dominates and the effective number of flavors decreases.
Since the matter passes through the high $\muB$ and low $T$ region
compared with high-energy heavy-ion collisions,
the reduction of the CP temperature $\TCP$ is essential
for the CP sweep during the BH formation.

%Chiral effective models of QCD have been extensively utilized
%to study the phase transition at finite $\mu$ region.
The Lagrangian density of the Polyakov loop extended quark meson (PQM) model,
as an example of chiral effective models, is given by
\begin{align}
{\cal L} =& \bar{q}\left[
i\gamma^\mu D_\mu
- g(\sigma + i\gamma_5\bm\tau\cdot\bm\pi)
- g_v\gamma^\mu(\omega_\mu + \bm\tau\cdot\bm{R}_\mu)
\right]q 
\nonumber\\
+& \frac{1}{2}(\partial_\mu\sigma)^2 +
\frac{1}{2}(\partial_\mu\bm\pi)^2 -U(\sigma,\bm\pi)
-{\cal U}_\ell(\ell,\bar{\ell},T)
\nonumber\\
-& \frac{1}{4} \omega_{\mu\nu}\omega^{\mu\nu}
- \frac{1}{4} \bm R_{\mu\nu} \cdot \bm R^{\mu\nu}
+ \frac{1}{2} m_v^2 (\omega_\mu\omega^\mu
	+ \bm R_\mu\cdot \bm R^\mu)
~.
\end{align}
The mesonic potential is
$U(\sigma,\bm\pi)=\lambda\left(\sigma^2+\bm\pi^2-v^2\right)^2/4-h\sigma$,
and $\omega_{\mu\nu}$ and $\bm R_{\mu\nu}$ are the field tensors of
the $\omega$ and $\rho$ mesons.
We use the Polyakov loop effective potential
$\mathcal{U}_\ell(\ell,\bar{\ell},T) = T^4[-a(T)\bar{\ell}\ell/2 + b(T)\log H(\ell,\bar{\ell})]$
where the Polyakov loop is defined as
$\ell = \text{Tr}[{\cal P}\exp(i\int_0^{1/T} d\tau A_4)]/N_c$.
In the mean field approximation, the thermodynamic potential is
obtained as
\begin{align}
\Omega_\mathrm{PQM} =& U_\sigma + \mathcal{U}_\ell
- \frac{g_v^2}{m_v^2}\left(\rho_u^2 + \rho_d^2\right)
% + \Omega_0 + \Omega_T
-2N_f N_c\int^\Lambda \frac{d^3\bm p}{(2\pi)^3} E_p
\nn\\
-& 2 T \sum_{f}\int\frac{d^3\bm p}{(2\pi)^3}
\left[ \log \mathcal{R}(E_p - \tilde{\mu}, \ell, \bar{\ell})
+ \log \mathcal{R}(E_p + \tilde{\mu}, \bar{\ell}, \ell)
\right]
\ ,\label{eq:Otemp}
\end{align}
where
$E_p=\sqrt{\bm p^2 + M^2}$,
$M=g\sigma+m_0$ is the constituent quark mass,
and 
$\tilde{\mu}_f = \mu \mp \delta\mu -2g_v^2\rho_f/m_v^2$
is the effective chemical potential
with $\mp=-, +$ for $f=u$ and $d$, respectively.
While the PQM model is renormalizable,
we adopt here the momentum cutoff $\Lambda$ for simplicity.

%**********************   III. Results   **************************************%

\begin{figure}[bt]
\begin{center}
\Psfig{6cm}{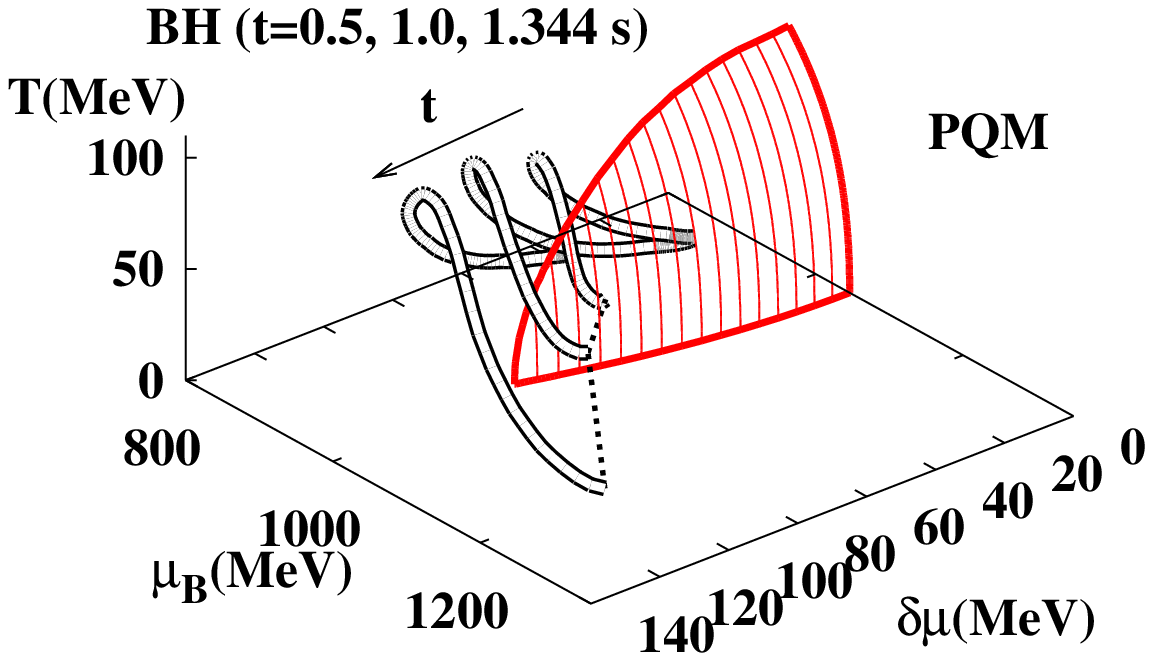}~\Psfig{6cm}{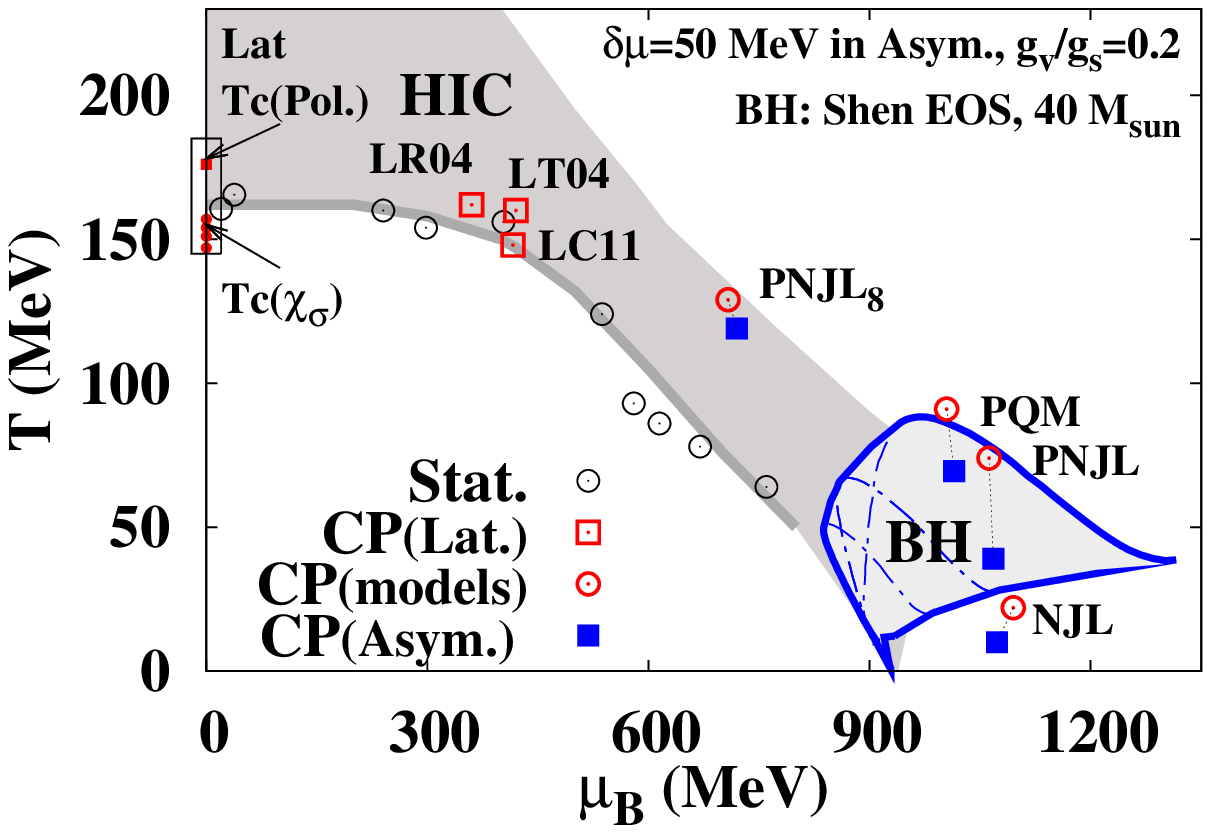}
\end{center}
\caption{
Left: First order phase boundary surface (solid lines) in $(T, \muB, \delta\mu)$ space
calculated with the PQM model are compared with the BH formation profile, 
(thermodynamical profile $(T,\muB,\delta\mu)$ during the BH formation)
at $t=0.5, 1.0, 1.344$ sec after the bounce (double lines).
Right: Predictions of the critical temperature and critical point locations
in comparison with the chemical freeze-out points
and the swept region during the black hole formation.
}\label{Fig:3D}
\end{figure}

We show the first order phase boundary of symmetric ($\delta\mu=0$)
and asymmetric ($\delta\mu\not= 0$) matter in PQM 
in the left panel of Fig.~\ref{Fig:3D}.
We find a trend that the first order phase boundary shrinks
at finite $\delta\mu$.
Transition temperature at a given baryon chemical potential
$\muB=3\mu$ decreases,
and the transition baryon chemical potential $\mu_c$ at $T=0$ also decreases.
We do not consider here the pion condensed phase,
because the $s$-wave $\pi N$ repulsion
would suppress the $s$-wave pion condensation~\cite{spion}.

The CP location is sensitive to $\delta\mu$.
Compared with the results in symmetric matter,
$\TCP$ becomes smaller at finite $\delta\mu$
and reaches zero at $\delta\mu=\delta\mu_c \simeq (50-80)~\MeV$.
CP is also sensitive to the model and parameters
as shown in the right panel of Fig.~\ref{Fig:3D}.
The Polyakov loop suppresses single quark excitations in the hadron phase,
then the transition temperature and thus the critical temperature
are shifted upwards in the P-NJL model.
The temporal component of the vector potential shifts the chemical potential
effectively, and consequently leads to an upward shift of $\mu_c$
by about 10-15 MeV at $g_v/g=0.2$.

We shall now compare the CP location and the phase boundary
with the BH formation profile. In the left panel of Fig.~\ref{Fig:3D},
we compare the phase boundaries and the BH formation profile in the PQM model.
During the BH formation,
the baryon chemical potential reaches around 1000,  1100 and 1300 MeV
in the central region of the proto-neutron star at
$t=0.5, 1.0$ and $1.344$ sec, respectively,
suggesting that quark matter would be formed
between $t=0.5$ and $1.0$ sec in the central region
% of the proto-neutron star
in most of the models considered here.
%From the comparison, 
We also find that the BH profile go through
the critical line in asymmetric matter,
{\em i.e.} the CP sweep takes place in PQM.
In other models, we find that 
there are three possible types in the transition to the quark matter
during the evolution of matter toward the BH formation;
the first order transition, the crossover transition, and the CP sweep,
where the BH formation profile goes below, above and through
the critical line in asymmetric matter.

The predicted CP locations
in lattice MC~\cite{LR04,LT04,LC11} and effective models~\cite{BHCP}
seem to be in the $(T,\muB)$ region probed
in heavy-ion collisions~\cite{Andronic,Stephanov}
or during the prompt black hole formation processes~\cite{BHCP}
as shown in the right panel of Fig.~\ref{Fig:3D}.
%We find that most of the predicted CP locations are
%in the accessible region in heavy-ion collisions or black hole formation.
Recent LQCD-MC predictions by using the reweighting (LR04)~\cite{LR04},
Taylor expansion (LT04)~\cite{LT04},
and canonical ensemble method (LC11)~\cite{LC11}
are consistent with each other, and suggest that the CP is accessible
in heavy-ion collisions.
It should be noted that some previous studies~\cite{LQCD-CP-old}
predicted larger $\muCP$, and there is also implication that
there is no chiral critical point
in the small $\mu$ region for $N_f=3$~\cite{Forcrand-Philipsen}.
By comparison, effective models generally predict the CP
in the lower $T$ and larger $\mu$ region, and many of them are accessible 
during the black hole formation.
Further studies are necessary to understand the difference
between the lattice MC and effective model results.

\section{Summary and Discussion}

We have discussed here two subjects on the QCD phase diagram at finite density.
One of them is the chiral and $\ZNc$ deconfinement transition boundaries
at finite $\mu$.
In the strong coupling lattice QCD with Polyakov loop effects~\cite{PSC},
the $\ZNc$ deconfinement boundary defined as the peak of $d\ell/dT$ is found
to deviate from the chiral transition boundary at finite $\mu$,
and it suggests the existence of the Polyakov loop suppressed
high density matter,
which may be interpreted as quarkyonic matter~\cite{Quarkyonic}.
This would be consistent with the lattice QCD Monte-Carlo simulation
results~\cite{BW2006},
which suggests that the $\ZNc$ deconfinement transition temperature $T_c(\mathrm{Pol.})$
would be meaningfully higher than the chiral transition temperature $T_c(\chi)$ at $\mu=0$.
The chemical potential effects should be stronger on $T_c(\chi)$
than on $T_c(\mathrm{Pol.})$,
then separation at $\mu=0$ ($T_c(\chi) > T_c(\mathrm{Pol.})$)
would be enhanced at finite $\mu$.
This contradicts to the results including additional chiral-Polyakov coupling,
or the chemical potential dependence of
the Polyakov loop potential~\cite{PNJL-B}.
In the strong coupling lattice QCD, the additional chiral-Polyakov coupling 
appears in the higher order terms of the large-dimensional ($1/d$) expansion,
and its effects should be studied.

The possibility to probe the QCD critical point
during black hole formation processes is discussed in the second 
part~\cite{BHCP}.
We have found that the critical point in isospin asymmetric matter
would be accessible in the black hole formation processes,
if the CP is in the low $T$ and high $\mu$ region
as predicted in chiral effective models.
It should be noted that we compare
the results of the CP location in chiral effective models
and the thermodynamical condition $(T,\muB)$ calculated with the hadronic EOS.
This comparison is relevant,
since the thermal trajectory should be the same
even if we use the combined EOS of quark and hadronic matter,
as long as the hadronic EOS is reproduced at low $T$ and $\muB$ in the combined EOS.
It is desired to examine the thermodynamical evolution in the combined EOS.
Another interesting point is the differences of the CP locations
in chiral effective models and in lattice MC simulations.
While lattice MC simulations have the sign problem and
there is also implications that there is no chiral critical point
in the low $\mu$ region~\cite{Forcrand-Philipsen},
recent results consistently suggest the CP may exist
in the low chemical potential region, $\muCP \simeq 500~\MeV$.
It would be necessary to understand these differences in order to pin down
the CP location in the phase diagram.

%\section*{Acknowledgement}

\medskip

%The author would like to thank Prof. A. Nakamura for useful discussions.
This work is supported in part by
Grant-in-Aid for Scientific Research from JSPS and MEXT
(Nos. 22-3314, 22540296),
the Grant-in-Aid for Scientific Research on Innovative Areas
from MEXT (No. 20105004),
the Global COE Program
"The Next Generation of Physics, Spun from Universality and Emergence",
and the Yukawa International Program for Quark-hadron Sciences (YIPQS).

\end{document}